\documentclass[pra,twocolumn,showpacs,amssymb]{revtex4}
\newcommand{\be}{\begin{equation}}
\newcommand{\ee}{\end{equation}}
\newcommand{\bea}{\begin{eqnarray}}
\newcommand{\eea}{\end{eqnarray}}
\usepackage{graphicx}
\usepackage{amsmath}

\begin{document}

\title{Superfluid transition in the attractive Hofstadter-Hubbard model}
\author{R. O. Umucal\i lar}
\author{M. Iskin}
\affiliation{Department of Physics, Ko\c{c} University, Rumelifeneri Yolu, 
34450 Sar\i yer, Istanbul, Turkey}

\date{\today}
\begin{abstract}
We consider a Fermi gas that is loaded onto a square optical lattice and 
subjected to a perpendicular artificial magnetic field, and determine its 
superfluid transition boundary by adopting a BCS-like mean-field approach 
in momentum space. The multi-band structure of the single-particle 
Hofstadter spectrum is taken explicitly into account while deriving a 
generalized pairing equation. We present the numerical solutions as 
functions of the artificial magnetic flux, interaction strength, Zeeman field, 
chemical potential, and temperature, with a special emphasis on the 
roles played by the density of single-particle states and center-of-mass 
momentum of Cooper pairs.
\end{abstract}

\pacs{03.75.Ss, 03.75.Hh, 64.70.Tg, 67.85.-d, 67.85.-Lm}

\maketitle

\section{Introduction}
\label{sec:intro}

Experiments with ultra-cold atomic systems have incessantly progressed 
in the past two decades or so, since the creation of the very first BEC with 
a dilute gas of bosonic atoms. Equipped with unique opportunities in 
controlling a wide-range of parameters, these systems have successfully 
been employed in not only testing numerous theoretical models developed 
in the condensed-matter literature but also studying new 
phenomena which do not have a direct analogue in other fields~\cite{cold atoms}. 
For instance, by changing the effective interaction strength between atoms 
through what is known as magnetic Feshbach resonances, the so-called 
BCS-BEC crossover has been experimentally realized with a superfluid 
(SF) Fermi gas, although such a phenomenon was originally explored 
from a theoretical perspective in the context of high-$T_c$ 
superconductors~\cite{BCS-BEC}. Similarly, the basic mechanism 
for unconventional pairings that has been experimentally realized 
with a population-imbalanced SF Fermi gas was originally proposed as a
mechanism for inhomogeneous superconductivity that is caused by the 
Zeeman-induced mismatch of the Fermi surfaces, long before the advent 
of atomic systems~\cite{FFLO,Koponen}. More recently, there has been 
a fervent activity in the cold-atom community to realize quantum-Hall-like 
effects with charge-neutral atoms through the use of artificial gauge fields 
and synthetic dimensions~\cite{Artificial}.

In this paper, we are interested in a merger of these topics, \textit{i.e.}, we 
study the SF transition of a Fermi gas that is loaded onto a square optical 
lattice and subjected to a perpendicular artificial magnetic field in the 
context of the so-called attractive Hofstadter-Hubbard model. 
Limited aspects of this problem were investigated both in momentum space 
within the BCS-like mean-field approach paying attention to single-particle 
degeneracies~\cite{Zhai, Maska}, and in real space within the Bogoliubov-de 
Gennes formalism including the possibility of imbalanced chemical and/or 
vector potentials~\cite{Iskin}. Here, we focus on determining the SF 
transition boundary as functions of the artificial magnetic flux, interaction 
strength, Zeeman field, chemical potential, and temperature. 
In comparison to the existing literature, we not only develop a better 
understanding of the pairing mechanism in momentum space but also 
locate the transition boundary more precisely within the adopted 
approximations. We also examine the roles played by the density of 
single-particle states and center-of-mass (CoM) momentum of 
Cooper pairs on the transition boundary, providing clear insights into the 
intriguing re-entrant superfluidity behavior found in the numerical solutions.
We trace the origin of this re-entrant behavior back to the strongly modified 
density of single-particle states in the presence of a magnetic flux. The 
magnetic flux splits the original band of the field-free case into several 
subbands~\cite{Hofstadter}. As a result of such a change in the band 
structure, the density of single-particle states becomes a non-monotonic 
function of energy, imposing a similar non-monotonic behavior on the 
phase boundaries.

The rest of the paper is organized as follows. In Sec.~\ref{sec:mft}, we first 
introduce the attractive Hofstadter-Hubbard model, and then obtain a 
self-consistent equation for the SF transition boundary by tackling the 
model Hamiltonian with a BCS-like mean-field approach in momentum space. 
Our numerical results are given in Sec.~\ref{sec:numerics}, where we 
present the phase boundaries in interaction strength-Zeeman field, 
interaction strength-chemical potential and temperature-chemical potential 
planes for a number of magnetic flux values. We conclude the paper with 
a brief summary in Sec.~\ref{sec:conc}. In addition, detailed derivations of 
the Hofstadter spectrum and generalized pairing equation are outlined,
respectively, in Appendices~\ref{sec:sps} and~\ref{sec:gpe}, and 
additional phase diagrams are included in Appendix ~\ref{sec:pd}.

\section{Mean-field Theory}
\label{sec:mft}

Our fundamental assumption is that the motion of a single particle in a 
tight-binding square optical lattice that is subjected to a perpendicular 
magnetic field is well-described by the famous Hofstadter model
\bea 
H_B = - t\sum_{\langle ij\rangle \sigma}\left(e^{i2\pi \phi_{ij}} 
c^\dag_{i\sigma}c_{j\sigma}+\text{h.c.}\right), 
\label{eq:Hofstadter Hamiltonian}
\eea
where $c^{\dagger}_{i\sigma}$ ($c_{i\sigma}$) creates (annihilates) a 
fermion with pseudo-spin $\sigma\equiv\{\uparrow,\downarrow\}$ at site $i$, 
\text{h.c.} is the Hermitian conjugate,
$t > 0$ is the hopping amplitude between nearest-neighbor sites 
$\langle ij\rangle$, and 
$
\phi_{ij} = (1/\phi_0) \int_{{\bf r}_j}^{{\bf r}_i} {\bf A}\cdot d{\bf r}
$ 
is the spin-independent phase factor the particle acquires while hopping 
from site $j$ to $i$. Here, $\phi_0 = \hbar/q_0$ is the effective magnetic-flux 
quantum with $q_0$ the effective charge, and ${\bf A} = (0, Bx)$ is 
the vector potential in the Landau gauge with $B$ the magnitude
of the effective magnetic field. Note that neither $q_0$ nor $B$ corresponds 
to a physical quantity by itself in atomic systems that are engineered to
simulate artificial gauge fields, but only their product is physically 
meaningful. When the particle traverses a loop encircling a unit cell of 
the lattice, its wave function acquires the Aharonov-Bohm phase factor 
$e^{i2\pi\alpha}$, where $\alpha = Ba^2/\phi_0$ is the flux quanta per 
unit cell with $a \to 1$ the lattice constant. As we outlined in 
Appendix~\ref{sec:sps} for completeness, when $\alpha$ is a rational 
fraction $p/q$ with $p$ and $q$ co-prime integers, the tight-binding 
$s$-band of the single-particle spectrum in the field-free case splits into 
$q$ subbands yielding the so-called Hofstadter butterfly which is a 
self-similar function of $\alpha$~\cite{Hofstadter}. The 
non-interacting Hamiltonian 
$
H_0 = H_B - \sum_{i \sigma} \mu_\sigma n_{i\sigma}
$ 
in the grand-canonical ensemble can equivalently be expressed as 
$
H_0 = H_B 
- \mu \sum_i (n_{i\uparrow} + n_{i\downarrow})
- h \sum_i (n_{i\uparrow} - n_{i\downarrow}),
$ 
where 
$
n_{i\sigma} = c^{\dagger}_{i\sigma}c_{i\sigma}
$ 
is the number operator, $\mu = (\mu_{\uparrow}+\mu_{\downarrow})/2$ is 
the average chemical potential, and $h = (\mu_{\uparrow}-\mu_{\downarrow})/2$ 
can be interpreted as an out-of-plane Zeeman field. 

We restrict ourselves to on-site atom-atom interactions that are described 
by the attractive Hubbard Hamiltonian
$
H_I = U \sum_{i} c^\dag_{i\uparrow}c^\dag_{i\downarrow}c_{i\downarrow}c_{i\uparrow},
$
where $U \le 0$. Adopting a BCS-like mean-field approximation for 
pairing, \textit{i.e.}, assuming that the fluctuations of the quadratic 
operators $c_{i\downarrow}c_{i\uparrow}$ are small in comparison
to their equilibrium values, we may decouple $H_I$ as
\be 
H_I \approx -\sum_{i}\left(\Delta_i c^\dag_{i\uparrow}c^\dag_{i\downarrow}
+\Delta^\ast_i c_{i\downarrow}c_{i\uparrow}+\frac{|\Delta_i|^2}{U}\right)
\label{eq:real space MF interaction},
\ee
where the complex order parameter
$
\Delta_i = U \langle c_{i\uparrow} c_{i\downarrow} \rangle
$
describes the on-site atom-atom correlations in thermal equilibrium 
as denoted by the thermal average $\langle \ldots \rangle$. 
The SF phase is characterized by $\Delta_i \neq 0$ at least for some $i$.
When $\Delta_i = 0$ for all $i$, the spin-$\sigma$ particles are either 
a normal Fermi gas or form a band insulator depending on their thermal 
average numbers determined by
$
N_{i\sigma} = \langle n_{i\sigma}\rangle.
$
Due to the Pauli exclusion principle, $N_{\sigma} = \sum_i N_{i\sigma}$ 
can at most be the total number of lattice sites $M = M_x M_y$, 
corresponding to a fully-occupied spectrum for any given $\alpha$.
However, when the number of fully-occupied magnetic subbands for 
spin-$\sigma$ particles is precisely an integer $s \leq q$ such that 
$N_{\sigma}/M = s/q$, or equivalently $\mu_\sigma$ is inside the 
corresponding single-particle energy gap, the particles form a band 
insulator. Otherwise, they are normal. 

A compact closed-form expression for the SF transition boundary can be 
obtained in momentum ($\mathbf{k}$) space where it is relatively easier to 
diagonalize $H_B$. For this purpose, we introduce the $\mathbf{k}$-space 
operators 
\bea 
c_{{\bf k}\beta\sigma} = \sqrt{\frac{q}{M_x M_y}} 
\sum\limits_{ s = 0}^{M_x/q}\sum\limits_{ i_y = 0}^{M_y}
c_{s\beta i_y\sigma}e^{-ik_x s q}e^{-i k_y i_y},
\label{eq:k-space c}
\eea
where $M_x$ and $M_y$ are, respectively, the number of lattice sites 
along the $x$ and $y$ directions with periodic boundary conditions 
in mind, and ${\bf k} = (k_x, k_y)$ is the momentum vector. 
Here, the real-space coordinate of site $i$ is expressed as 
${\bf r}_i = (i_x, i_y)$, where $i_x = s q+\beta$ with $s = 0,\ldots,M_x/q$ 
denoting the location of the enlarged ($q \times 1$) unit cell in the 
lattice and $\beta = 0,\ldots,q-1$ denoting a particular site inside the 
enlarged unit cell. Since such choice of an enlarged unit cell restores 
the translational symmetry of the original lattice for the particular 
Landau gauge of interest, it allows us to retain the Bloch description 
of the eigenstates with a reduced (magnetic) Brillouin zone (MBZ): 
$k_x \in [-\pi/q,\pi/q)$ and $k_y \in [-\pi,\pi)$~\cite{Hofstadter,Kohmoto}. 

Using Eq.~\eqref{eq:k-space c} in Eq.~\eqref{eq:Hofstadter Hamiltonian},
we obtain
$
H_B = \sum_{{\bf k}\sigma}\sum_{\alpha\beta} c^{\dagger}_{{\bf k}\alpha\sigma}
H_{{\bf k}\sigma}^{\alpha\beta}c_{{\bf k}\beta\sigma},
$ 
where the matrix elements $H_{{\bf k}\sigma}^{\alpha\beta}$ are explicitly 
given in Appendix~\ref{sec:sps}. Diagonalization of this $q\times q$ 
matrix yields $q$ eigenvalues $\varepsilon_{{\bf k}n\sigma}$ for a given 
$\mathbf{k}$ with $n = 1,\ldots,q$ corresponding to $q$ subbands that
split from the original field-free band. Note that the single-particle spectrum 
$
\varepsilon_{{\bf k}n\sigma}=\varepsilon_{{\bf k}n}
$
is spin-independent.
Using the band operators $d_{{\bf k}n\sigma}$ defined through the relation 
$
c_{{\bf k}\beta\sigma} = \sum_n g^n_\beta({\bf k})d_{{\bf k}n\sigma},
$ 
where $g^{n}_{\beta}({\bf k})$ is the $\beta$th component of the $n$th 
eigenvector of the single-particle problem with energy $\varepsilon_{{\bf k}n}$, 
and including $\mu_\sigma$, the non-interacting Hamiltonian finally 
reads as
\bea
H_0=\sum_{{\bf k}n\sigma}\epsilon_{{\bf
k}n\sigma}d^\dag_{{\bf k}n\sigma} d_{{\bf k}n\sigma},
\label{eq:H0}
\eea
where $\epsilon_{{\bf k}n\sigma}=\varepsilon_{{\bf k}n}-\mu_{\sigma}$ 
with ${\bf k}$ restricted to the first MBZ. Following a similar procedure, 
the Hamiltonian \eqref{eq:real space MF interaction} can be written 
in $\mathbf{k}$-space as
\bea
H_I &=& -\sum\limits_{l\beta}
\left\{\sum\limits_{n n^\prime{\bf k}}\left[\Delta^l_{\beta}
g^{n *}_{\beta}({\bf k}^{l}_{+})g^{n^\prime *}_{\beta}({\bf k}^{l}_{-}) 
\right.\right. \nonumber\\ 
&& \left.\left.
\times d^\dag_{{\bf k}^{l}_{+}n\uparrow}d^\dag_{{\bf k}^{l}_{-}n^\prime\downarrow}
+\!\text{h.c.}\right]\!+\!\frac{M}{qU}|\Delta^l_\beta|^2\right\},
\label{eq:interaction}
\eea
where the complex coefficients 
$
\Delta^{l}_\beta = -(qU/M)\sum_{n n^\prime{\bf k}}
g^n_\beta({\bf k}^{l}_{+})g^{n^\prime}_\beta({\bf k}^{l}_{-})\langle 
d_{{\bf k}^{l}_{-}n^\prime\downarrow}d_{{\bf k}^{l}_{+}n\uparrow}\rangle\label{eq:Delta}
$
are defined in such a way that
$
\Delta_i = \sum_{l} \Delta^{l}_\beta e^{i (Q_{lx} s + Q_{ly} i_y)}.
$
Here, ${\bf k}^{l}_{\pm} = \pm{\bf k}+{\bf Q}_l/2$ with
${\bf Q}_l = (Q_{lx}, Q_{ly})$ the CoM momentum of Cooper pairs. 
While all possible CoM momenta must in principle be allowed in the 
calculations, such a task is not numerically tractable for arbitrary $\alpha$. 
For this reason, we limit our numerical calculations mainly to BCS-like pairings
and consider a finite set ${\bf Q}_l=(0, 2\pi l p/q)$ with $l =0,\dots,q-1$.
Finite CoM pairing ${\bf Q}_l=(0, 2\pi l p/q)$ with $l \neq 0$, 
in addition to the usual BCS pairing with ${\bf Q}_l=(0,0)$, needs 
to be taken into account due to the degeneracy of the single-particle 
energies in any given band $n$ for momenta ${\bf k}$ and 
${\bf k}+{\bf Q}_l$, \textit{i.e.},
$
\varepsilon_{{\bf k}n}=\varepsilon_{{\bf k}+{\bf Q}_l,n}$~\cite{Zhai}. 
In the absence of a Zeeman field, we do not expect this limitation to 
a finite set of CoM momenta to have 
any effect on the SF transition boundary of interest here, even though 
the SF order parameter may slightly be affected by it deeper into the 
SF region. In the presence of imbalanced populations, while we expect 
this limitation to have some but minor influence on the SF transition 
boundary, we note that extending the calculation to FFLO-like pairings 
(\textit{e.g.}, by including additional CoM momenta to take explicitly the 
Zeeman-induced mismatch of the Fermi surfaces into account) 
may lead to a dramatic improvement in case a more accurate real-space 
description of the SF order parameter is desired.

Under these approximations, and noting that all of the coefficients 
$\Delta^{l}_\beta$ are expected to be small in the vicinity of the SF transition
boundary, $H_I$ may be treated as a perturbative correction to $H_0$. 
Using the first-order perturbation theory outlined in Appendix~\ref{sec:gpe}, 
we obtain a compact expression for the generalized pairing equation
\bea
\Delta^{l}_\beta = &-&\frac{qU}{M} \sum_{n n^{\prime} {\bf k} \beta^{\prime}}
g^{n}_\beta({\bf k}^{l}_{+})g^{n^\prime}_{\beta}({\bf k}^{l}_{-})
g^{n *}_{\beta^\prime}({\bf k}^{l}_{+})g^{n^\prime *}_{\beta^\prime}({\bf k}^{l}_{-})
\nonumber\\ 
&& \times 
\Delta^{l}_{\beta^\prime}
\frac{1-f(\epsilon_{{\bf k}^{l}_{+}n\uparrow})
-f(\epsilon_{{\bf k}^{l}_{-}n^\prime\downarrow})}
{\epsilon_{{\bf k}^{l}_{+}n\uparrow}+\epsilon_{{\bf k}^{l}_{-}n^\prime\downarrow}},
\label{eq:Gap Equation}
\eea
which determines the SF transition boundary for a given ${\bf Q}_l$. 
Here, $f(x) = 1/[e^{x/(k_B T)} + 1]$ is the usual Fermi-Dirac 
distribution function with $k_B$ the Boltzmann constant and $T$ the 
temperature. Note that Eq.~\eqref{eq:Gap Equation} has to be 
supplied simultaneously with the number equations
$
N_{\sigma} = \sum_{n{\bf k}}f(\epsilon_{n{\bf k}\sigma}),
$ 
forming a complete set of self-consistency equations for 
$\Delta^l = (\Delta^{l}_0, \ldots, \Delta^{l}_{q-1})$ and $\mu_\sigma$.
It is convenient to express Eq.~\eqref{eq:Gap Equation} in the form 
a matrix-eigenvalue equation, where 
$
\Delta^{l}_\beta = \sum_{\beta^{\prime}}
\mathbb{M}^l_{\beta\beta^\prime}\Delta^{l}_{\beta^\prime}
$ 
or equivalently $\Delta^{l} = \mathbb{M}^l\Delta^{l}$, from which
the condition for a nontrivial yet arbitrarily small $\Delta^{l}$ solution is 
determined by setting ${\rm det}(\mathbb{I}-\mathbb{M}^l) = 0$ 
with $\mathbb{I}$ the identity matrix. In case of multiple solutions 
for $U_c^l$ and $T_c^l$ that are allowed by the determinant condition, 
we ultimately identify $U_c = \max \{ U_c^l \}$ (or equivalently 
$|U_c| = \min \{ |U_c^l| \}$) as the critical interaction strength and 
$T_c = \max \{T_c^l\}$ as the critical temperature of the system.

These critical parameters depend sensitively on $\alpha$ directly through 
the resultant density of single-particle states 
$
D(\varepsilon) = d{\cal N}(\varepsilon)/d\varepsilon,
$ 
where ${\cal N}(\varepsilon)$ is the number of states per 
unit area with energy smaller than $\varepsilon$, and it can be calculated by simply 
counting the number of states $\Delta {\cal N}(\varepsilon)$ contained in a 
small interval of energy $[\varepsilon,\varepsilon+\Delta\varepsilon]$ with 
fixed $\Delta\varepsilon$. Since the spectrum is symmetric around 
$\varepsilon = 0$, we only consider $\varepsilon \geq 0$ as discussed below.

\section{Numerical Results}
\label{sec:numerics}

\begin{figure}[htbp]
\includegraphics[scale=0.45]{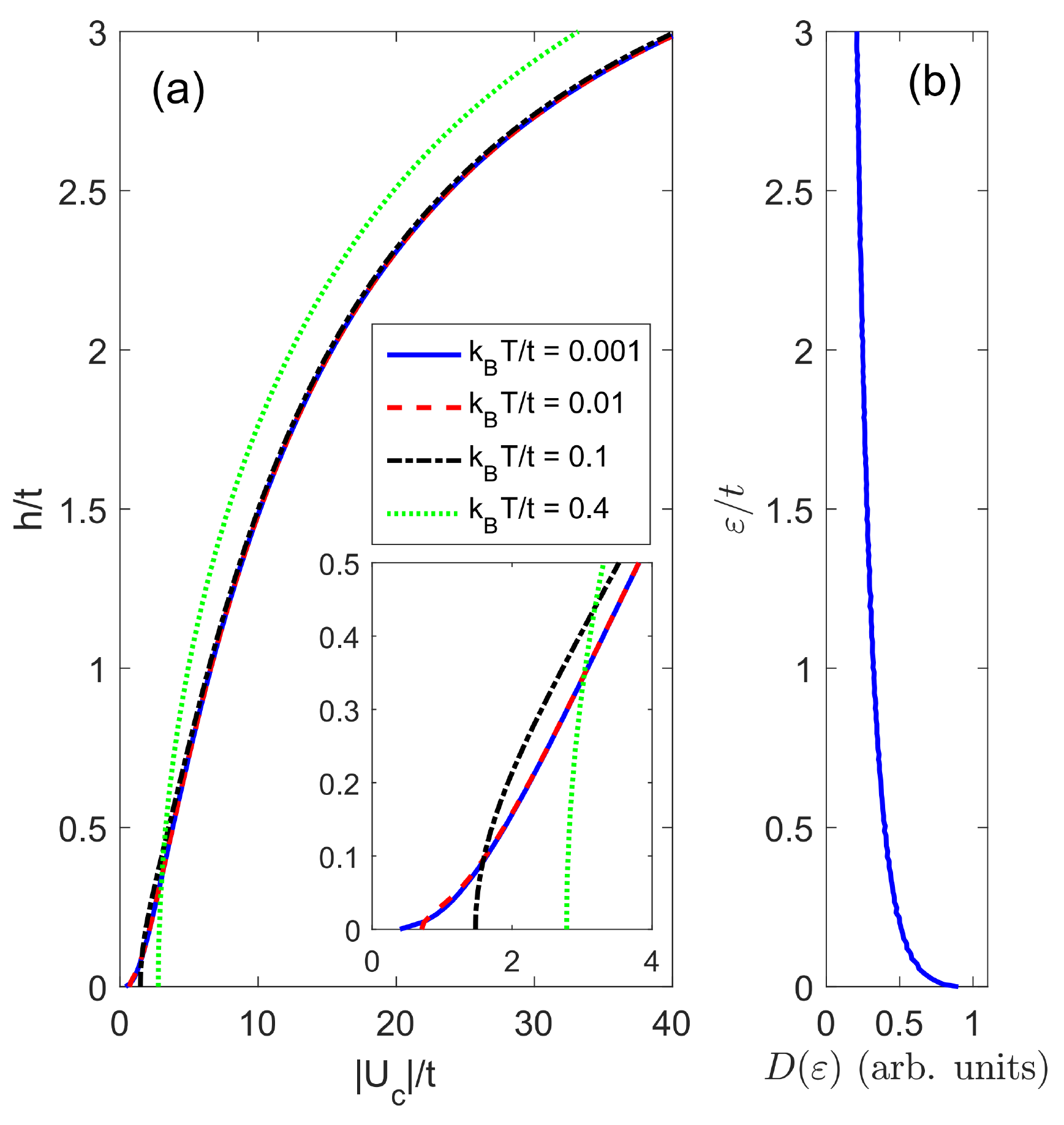}
\caption{(Color online) 
(a) Critical interaction strength $|U_c|/t$ versus the Zeeman field $h/t$ 
with $\alpha = 1/1$ and $\mu = 0$ for various temperatures $T$. 
The inset is a close-up for $h \leq 0.5t$. $|U_c|/t$ at $h = 0$ increases 
with $T$. (b) Density of states $D(\varepsilon)$ in arbitrary units. 
\label{p1q1mu0gcvsh}
}
\end{figure}

First of all, in the absence of a magnetic field ($B \to 0$ or $\alpha \to 0$), 
which can equivalently be accounted for by taking $p = q = 1$ and 
$l=\beta = 0$, the determinant condition reduces to the usual expression
$
-M/U_c = \sum_{{\bf k}}[1-f(\epsilon_{{\bf k}\uparrow})-f(\epsilon_{-{\bf k}\downarrow}]/(\epsilon_{{\bf k}\uparrow}+\epsilon_{-{\bf k}\downarrow}), 
$
where 
$
\epsilon_{\pm{\bf k}\sigma} = -2t(\cos k_x+\cos k_y)-\mu_{\sigma}
$ 
is the usual tight-binding spectrum shifted by the chemical potential. 
We recall that FFLO-like pairings~\cite{Koponen} are not considered in 
this work for the simplicity of the followup discussion. 
In Fig.~\ref{p1q1mu0gcvsh}(a), we show that $|U_c|$ is a monotonously
increasing function of the Zeeman field $h$, which follows closely the 
monotonous decrease of $D(\varepsilon)$ with increasing 
$\varepsilon$ ($\geq 0$) that is presented in Fig.~\ref{p1q1mu0gcvsh}(b).
This is simply because, since $h \ne 0$ changes the effective chemical 
potentials for spin-$\uparrow$ and -$\downarrow$ particles as 
$
\mu_{\uparrow,\downarrow} = \mu \pm h,
$ 
it directly effects the available number of states near the Fermi surface 
involved in pairing. When $D(\varepsilon)$ gets lower (higher), the 
formation of Cooper pairs is facilitated with a relatively large (small) $U_c$, 
which is a generic observation valid also in the presence of a magnetic field.

\begin{figure}[htbp]
\includegraphics[scale=0.45]{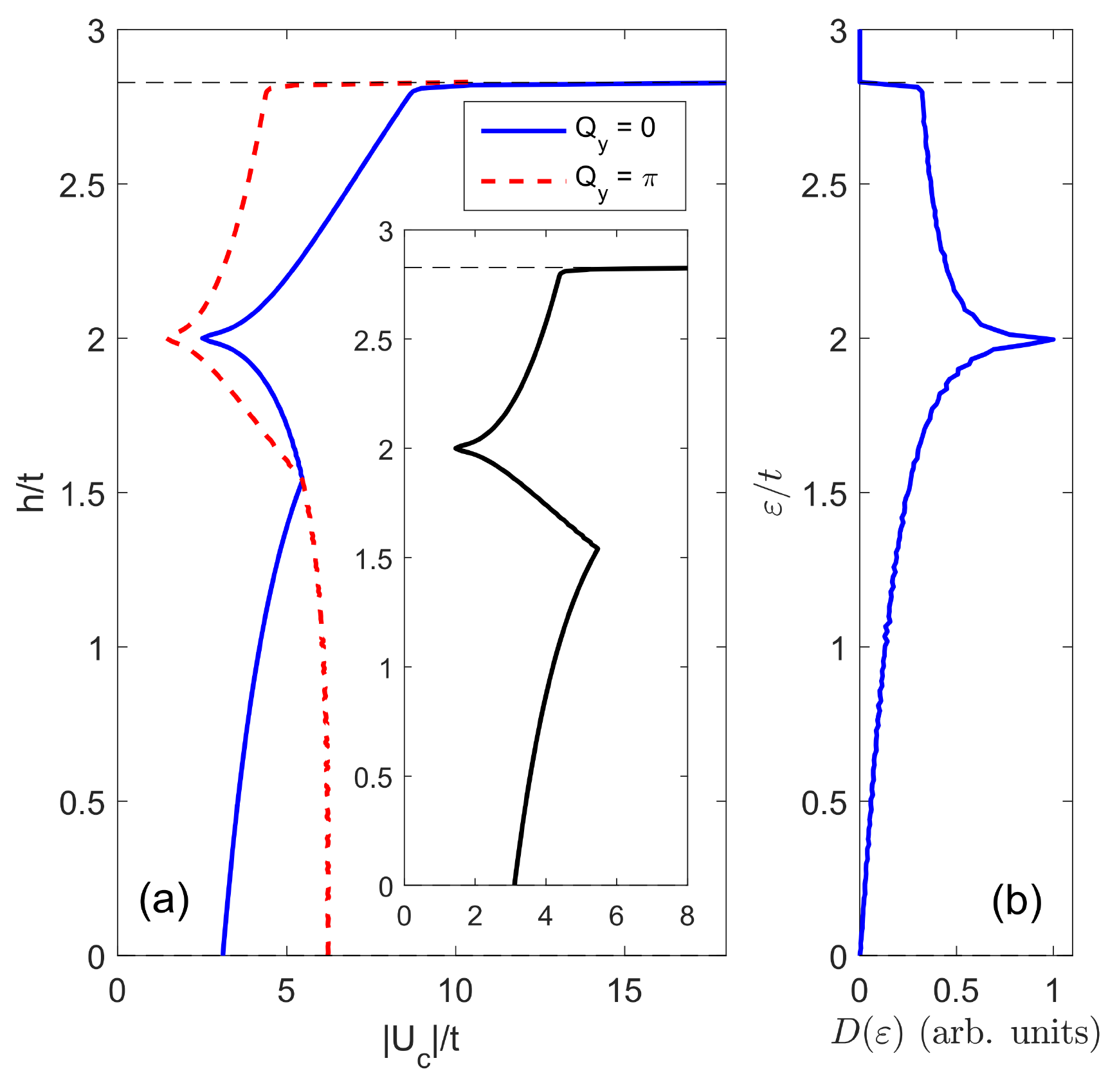}
\caption{(Color online) 
(a) Critical interaction strength $|U_c|/t$ versus 
the Zeeman field $h/t$ with $\alpha = 1/2$, $\mu = 0$, and $k_BT = 0.005t$. 
Solid blue curve is for ${\bf Q} = (0,0)$ and red dashed curve is for 
${\bf Q} = (0,\pi)$. The curve in the inset traces the minimum value 
of the two curves at each $h/t$. 
(b) Density of states $D(\varepsilon)$ in arbitrary units. 
Horizontal dashed lines show the band edges including $\varepsilon = 0$.
\label{p1q2mu0gcvsh}
}
\end{figure}

As the first example of a case with non-vanishing magnetic flux, we 
consider $\alpha = 1/2$ and set $\mu = 0$ for simplicity. The 
original field-free band splits into two bands that are touching each 
other at $\varepsilon = 0$. The singular peak of $D(\varepsilon)$ that 
is seen in Fig.~\ref{p1q2mu0gcvsh}(b) at $\varepsilon = 2t$ is due 
to a van Hove singularity, and it is directly reflected as a dip in $|U_c|$ 
precisely at $h = 2t$ that is shown in Fig.~\ref{p1q2mu0gcvsh}(a). 
More importantly, the figure inset illustrates that $U_c$ is determined 
either by ${\bf Q}_0 = (0,0)$ or ${\bf Q}_1 = (0,\pi)$ depending on 
the particular value of $h$. 

\begin{figure}[htbp]
\includegraphics[scale=0.45] {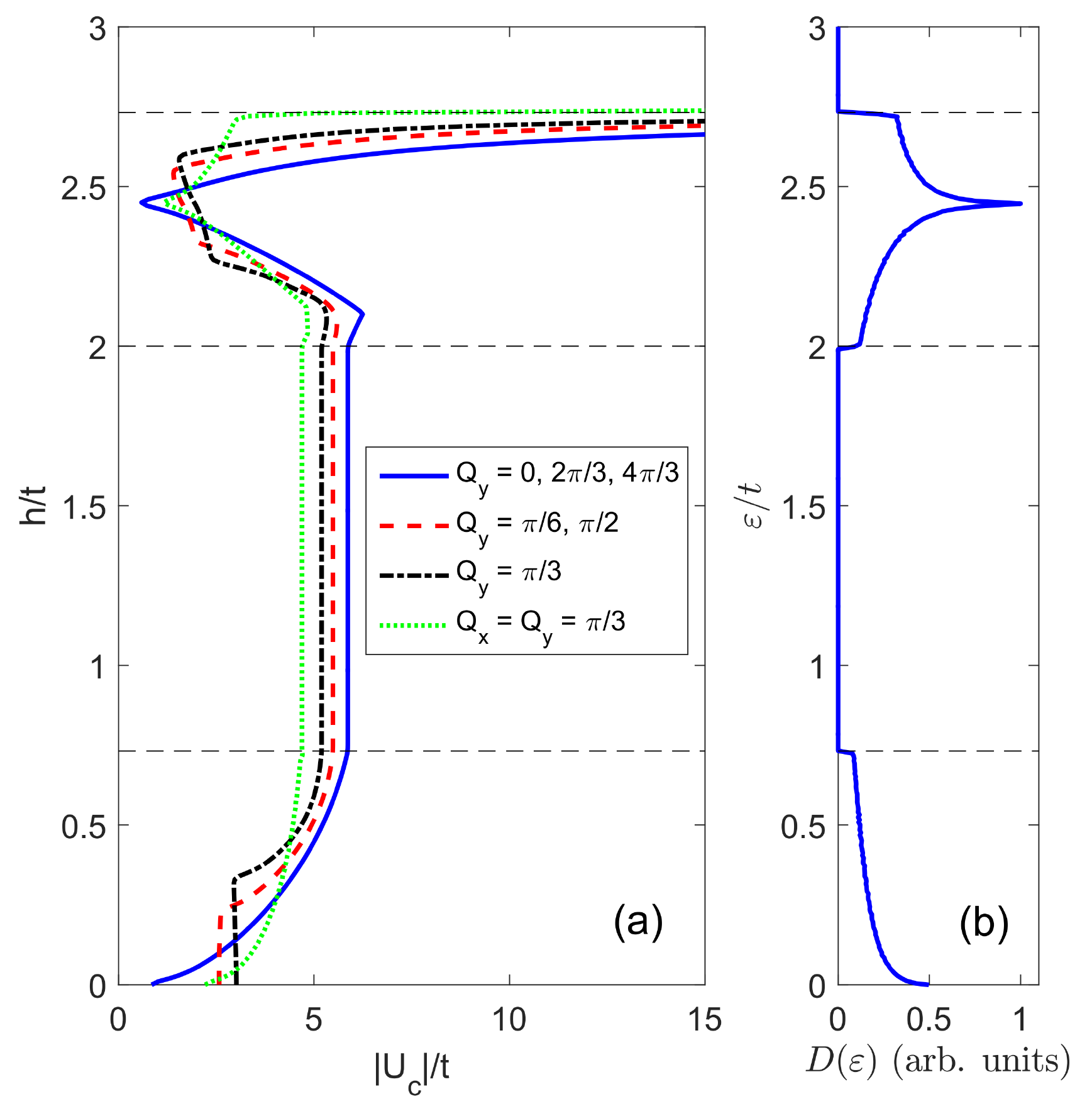}
\caption{(Color online) 
(a) Critical interaction strength $|U_c|/t$ versus the Zeeman field $h/t$ 
with $\alpha = 1/3$, $\mu = 0$, and $k_BT = 0.005t$. 
Solid blue curve is the phase boundary obtained for ${\bf Q}_l = (0, 2\pi l/3)$, 
with $l = 0,1,2$. Dashed red curve is for ${\bf Q} = (0,\pi/6)$ and 
${\bf Q} = (0,\pi/2)$, dash-dotted black curve is for ${\bf Q} = (0,\pi/3)$, 
and dotted green curve is for ${\bf Q} = (\pi/3,\pi/3)$. 
(b) Density of states $D(\varepsilon)$ in arbitrary units. 
Horizontal dashed lines show the band edges. 
\label{p1q3mu0gcvsh}
}
\end{figure}

As a second example shown in Fig.~\ref{p1q3mu0gcvsh}, we consider
$\alpha=1/3$ and again set $\mu = 0$. While there are three subbands 
in the spectrum, only the highest band and half of the middle band 
are seen in Fig.~\ref{p1q3mu0gcvsh}(b) since $D(\varepsilon)$ is
restricted to $\varepsilon \geq 0$. Figure~\ref{p1q3mu0gcvsh}(a) shows 
that $U_c^l$ are degenerate functions of $h$ for ${\bf Q}_l = (0, 2\pi l/3)$ 
with $l = 0,1$ and $2$, and the calculated dips correspond again to 
the peaks of $D(\varepsilon)$. 
It would be curious to check whether the set ${\bf Q}_l$ yields 
degenerate solutions for any $\alpha$ with odd denominators, 
\textit{e.g.}, see Appendix~\ref{sec:pd} for $\alpha = 1/5$. 
When $\mu_\uparrow = -\mu_\downarrow = h$ is inside the 
single-particle band gap, \textit{i.e.}, between $0.73t$ and $2t$, 
$|U_c|$ remains constant until new pairing possibilities appear as 
$\mu_\sigma$ crosses over to the upper/lower bands, leading to the 
observed re-entrant superfluidity behavior. Note in the gapped region 
that the ground state of the system is a band insulator with fillings 
$N_{\uparrow}/M = 2/3$ and $N_{\downarrow}/M = 1/3$ for $|U| < |U_c|$.

In Fig.~\ref{p1q3mu0gcvsh}(a), we also present the transition boundary
for three addional values of CoM momenta, namely ${\bf Q} = (0,\pi/6)$, 
${\bf Q} = (0,\pi/3)$, and ${\bf Q} = (0,\pi/2)$. While $|U_c|$ is smaller 
for our original set ${\bf Q}_l = (0, 2\pi l/3)$ with $l = 0, 1$ and $2$ 
near the peaks of $D(\varepsilon)$, these additional CoM momenta 
lead in general to close but lower $|U_c|$. Although we do not 
systematically study the dependence of $U_c$ on the additional 
CoM momentum for a given $\alpha$ and $h$, here is the plausibility 
argument for this observation. 

\begin{figure}[htbp]
\includegraphics[scale=0.45]{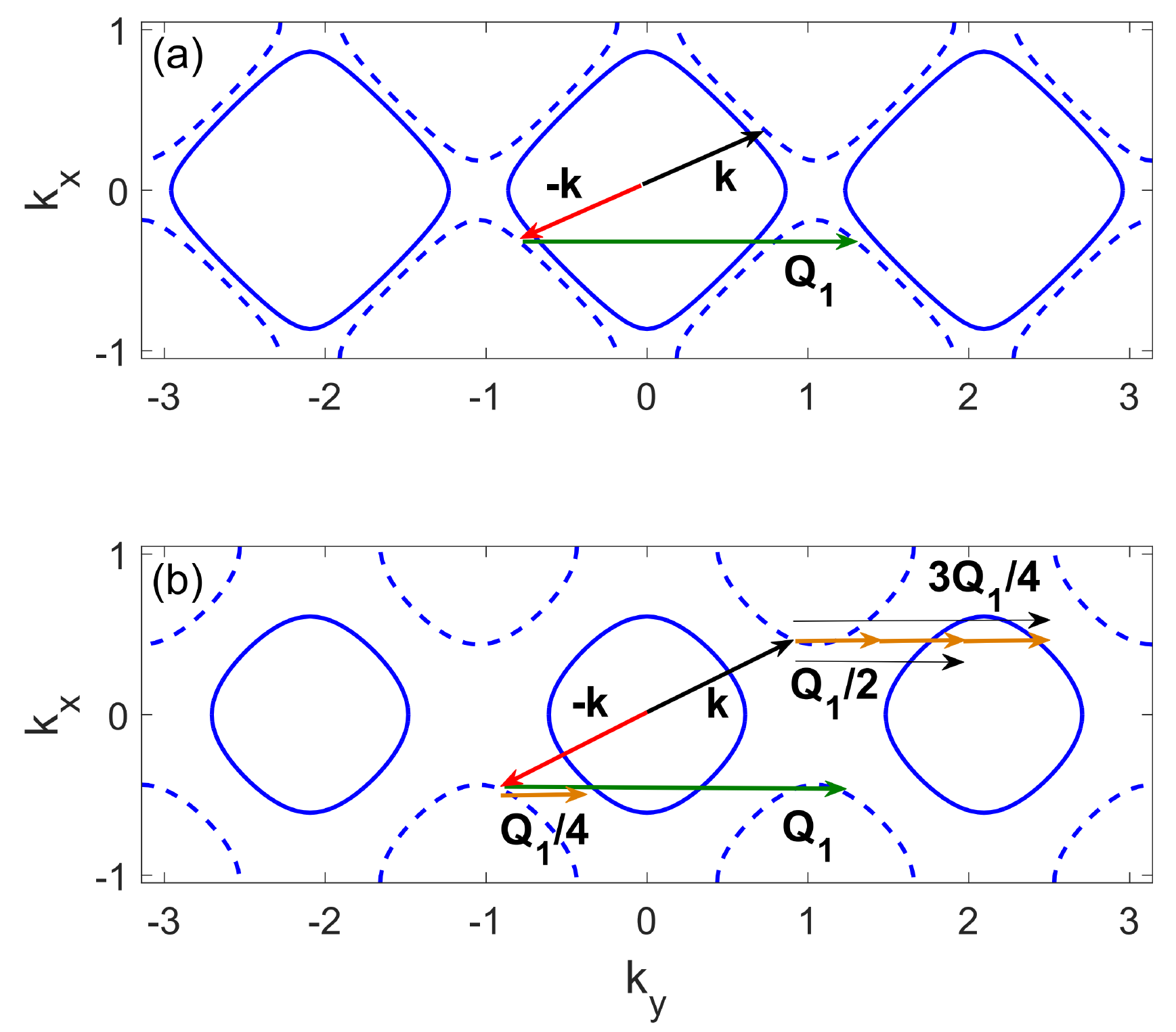}
\caption{(Color online) 
(a) The most plausible pairing is sketched between a spin-$\uparrow$ 
particle with momentum ${\bf -k}+{\bf Q}_1$ and a spin-$\downarrow$ 
particle with momentum ${\bf k}$ for $\alpha = 1/3$ and $\mu = 0$. 
Fermi surfaces $\mu_\uparrow = 0.05 t$ and $\mu_\downarrow = -0.05t$
are shown by solid and dashed lines, respectively.  
(b) Similar sketch for $h = 0.25t$ shows that the CoM momenta 
${\bf Q_1}/4, {\bf Q_1}/2$, and  $3{\bf Q}_1/4$ give better matching 
of the Fermi surfaces in comparison to the original ${\bf Q}_1$.
\label{p1q3momenta}
}
\end{figure}

In Fig.~\ref{p1q3momenta}, we set $\alpha = 1/3$ and $\mu = 0$, 
and sketch the Fermi surfaces of spin-$\uparrow$ (solid curves) 
and spin-$\downarrow$ (dashed curves) particles for two $h$ values, 
showing a number of pairing possibilities inside the middle 
band of the spectrum.
For small $h = 0.05t$, Fig.~\ref{p1q3momenta}(a) shows that a 
spin-$\downarrow$ particle with momentum ${\bf k}$ can be 
easily coupled to a spin-$\uparrow$ particle with momentum 
${\bf -k}+{\bf Q}_1$. Note that even though it is possible to find an 
arbitrary ${\bf Q}$ for a given ${\bf k}$ with the property of carrying 
${\bf -k}$ close to a solid curve, ${\bf Q}_1$ and its integer multiples 
have this property for all ${\bf k}$. Therefore, we expect such 
pairings to be enhanced over other types of pairing. 
For a larger $h = 0.25t$, Fig.~\ref{p1q3momenta}(b) clearly shows 
that the pairing of a spin-$\downarrow$ particle with momentum 
${\bf k}$ and a spin-$\uparrow$ particle with momentum ${\bf -k}+{\bf Q}_1$
is energetically much harder than the previous case, instead of 
which pairings with ${\bf Q_1}/4, {\bf Q_1}/2$ and $3{\bf Q}_1/4$ 
are relatively easier with a better match of the Fermi surfaces.
Furthermore, relaxing the condition on the vanishing $x$-component 
of the CoM momenta, \textit{e.g.}, ${\bf Q}_D = (\pi/3,\pi/3)$, allows 
for a nesting vector with perfect overlap between the Fermi surfaces,
despite an energy gap. Hence, we expect such CoM momentum 
vectors to be optimal when they are relevant. For example,
Fig.~\ref{p1q3mu0gcvsh}(a) shows that not only ${\bf Q}_D$ leads to 
the lowest $|U_c|$ when $h$ is inside a band gap but also its 
result may deviate significantly from other possibilities when 
$\mu_{\uparrow, \downarrow}$ is close to the highest/lowest band edge. 
A comprehensive analysis of such generalized pairing schemes is 
beyond the scope of this work, and we defer it to a future one.

\begin{figure}[htbp]
\includegraphics[scale=0.45] {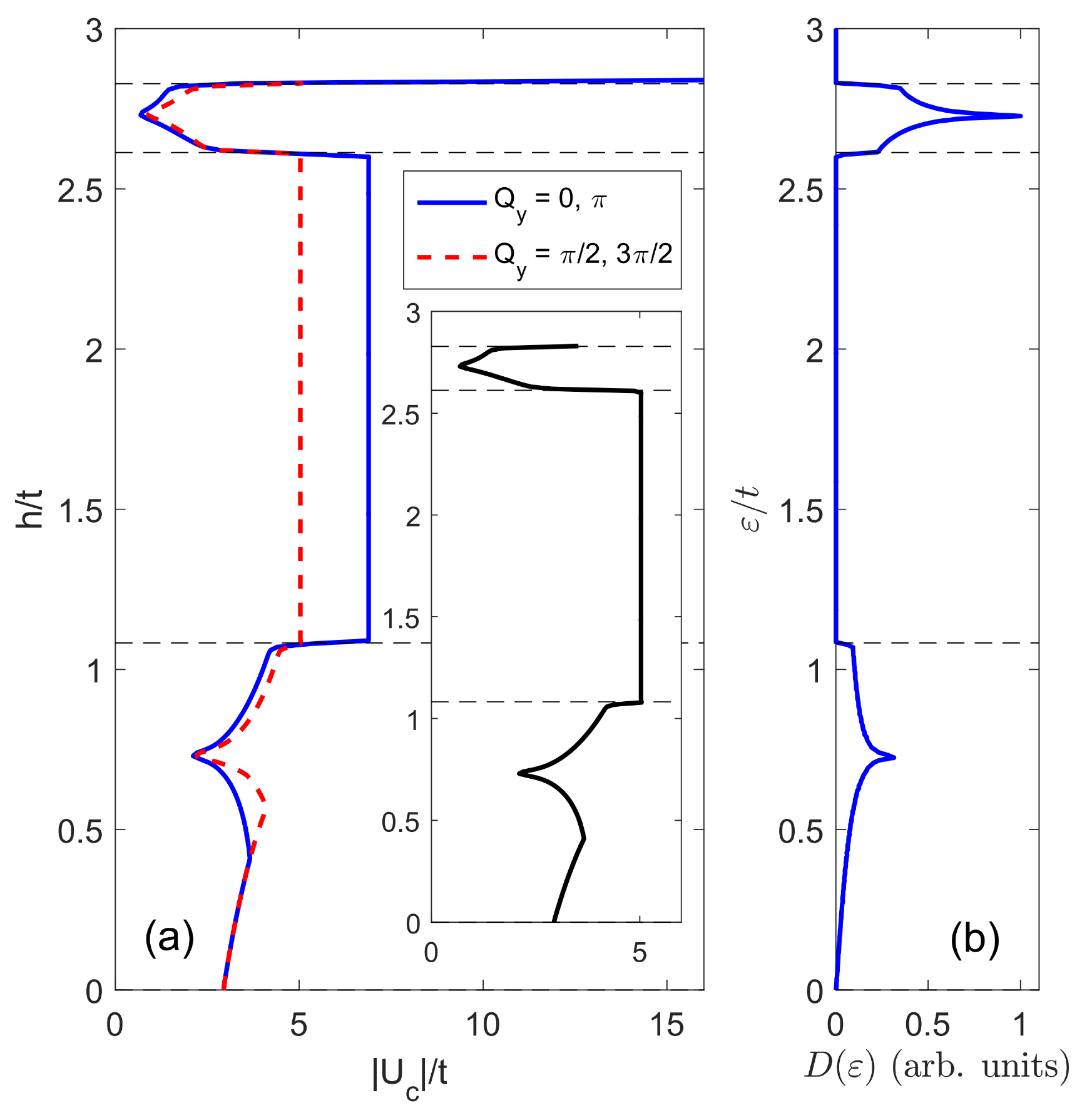}
\caption{(Color online) (a) 
Critical interaction strength $|U_c|/t$ versus the Zeeman field $h/t$ 
with $\alpha = 1/4$, $\mu = 0$ and $k_BT = 0.005t$. 
Solid blue curve is for ${\bf Q} = (0,0)$ and ${\bf Q} = (0,\pi)$; 
red dashed curve is for ${\bf Q} = (0,\pi/2)$ and ${\bf Q} = (0,3\pi/2)$. 
Inset shows the minimum value of the two curves at each $h/t$. 
(b) Density of states $D(\varepsilon)$ in arbitrary units. 
Horizontal dashed lines show the band edges including $\varepsilon = 0$.
\label{p1q4mu0gcvsh}
}
\end{figure}

The third example is shown in Fig.~\ref{p1q4mu0gcvsh}, where we 
consider $\alpha = 1/4$ and set $\mu = 0$. The original field-free band 
splits into a total of four bands, two of which are touching each other at 
$\varepsilon = 0$. Figure~\ref{p1q4mu0gcvsh}(a) shows that 
$U_c^l$ solutions corresponding to the $l = (0,2)$ and $l = (1,3)$ 
CoM momentum ${\bf Q}_l = (0, 2\pi l/4)$ are degenerate functions 
of $h$, and it would be curious to check whether such a grouping
of solutions is possible for any $\alpha$ with even denominators, 
\textit{e.g.}, see Appendix~\ref{sec:pd} for $\alpha = 1/6$. 
In accordance with the previous discussion, while $|U_c|$ again 
remains a constant when $\mu_\sigma = \pm h$ are inside the 
energy gap between the two highest single-particle bands, its ground 
state is a band insulator with fillings $N_{\uparrow}/M = 3/4$ and 
$N_{\downarrow}/M = 1/4$ for $|U| < |U_c|$.

\begin{figure}[htbp]
\includegraphics[scale=0.4]{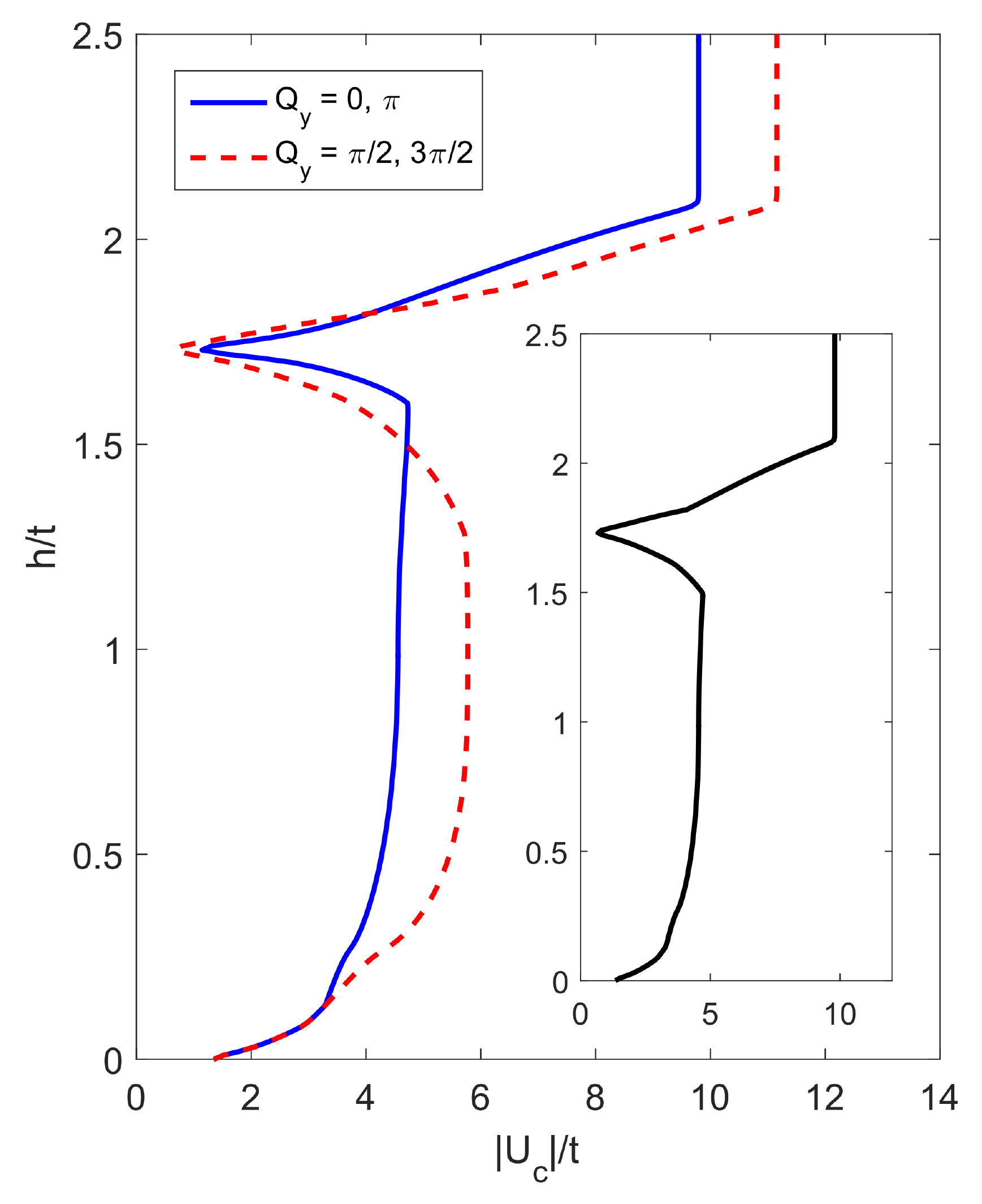}
\caption{(Color online) 
(a) Critical interaction strength $|U_c|/t$ versus the Zeeman field $h/t$ 
with $\alpha = 1/4$, $\mu = -t$ and $k_BT = 0.005t$. 
Solid blue curve is for ${\bf Q} = (0,0)$ and ${\bf Q} = (0,\pi)$; 
red dashed curve is for ${\bf Q} = (0,\pi/2)$ and ${\bf Q} = (0,3\pi/2)$. 
Inset shows the minimum value of the two curves at each $h/t$.
\label{p1q4mum1gcvsh}
}
\end{figure}

For completeness, next we again consider $\alpha =1/4$ but analyze 
the effects of $\mu \ne 0$ by setting $\mu = -t$. Since $\mu_{\sigma}$ 
is lowered by $-t$, Fig.~\ref{p1q4mum1gcvsh}(a) shows that $|U_c|$ 
has a single dip at around $h \simeq 1.7t$ within the range of $h$ 
presented. This peak corresponds to the enhanced pairing between a 
spin-$\uparrow$ particle with $\mu_{\uparrow} \simeq 0.7t$ from the 
middle band of the spectrum and a spin-$\downarrow$ particle with 
$\mu_{\downarrow} \simeq -2.7t$ from the lowest band. 
Similar to the $\mu = 0$ case, we again see that $U_c^l$ solutions 
corresponding to the $l = (0,2)$ and $l = (1,3)$ CoM momentum 
${\bf Q}_l = (0, 2\pi l/4)$ are degenerate functions of $h$.

\begin{figure}[htbp]
\includegraphics[scale=0.45]{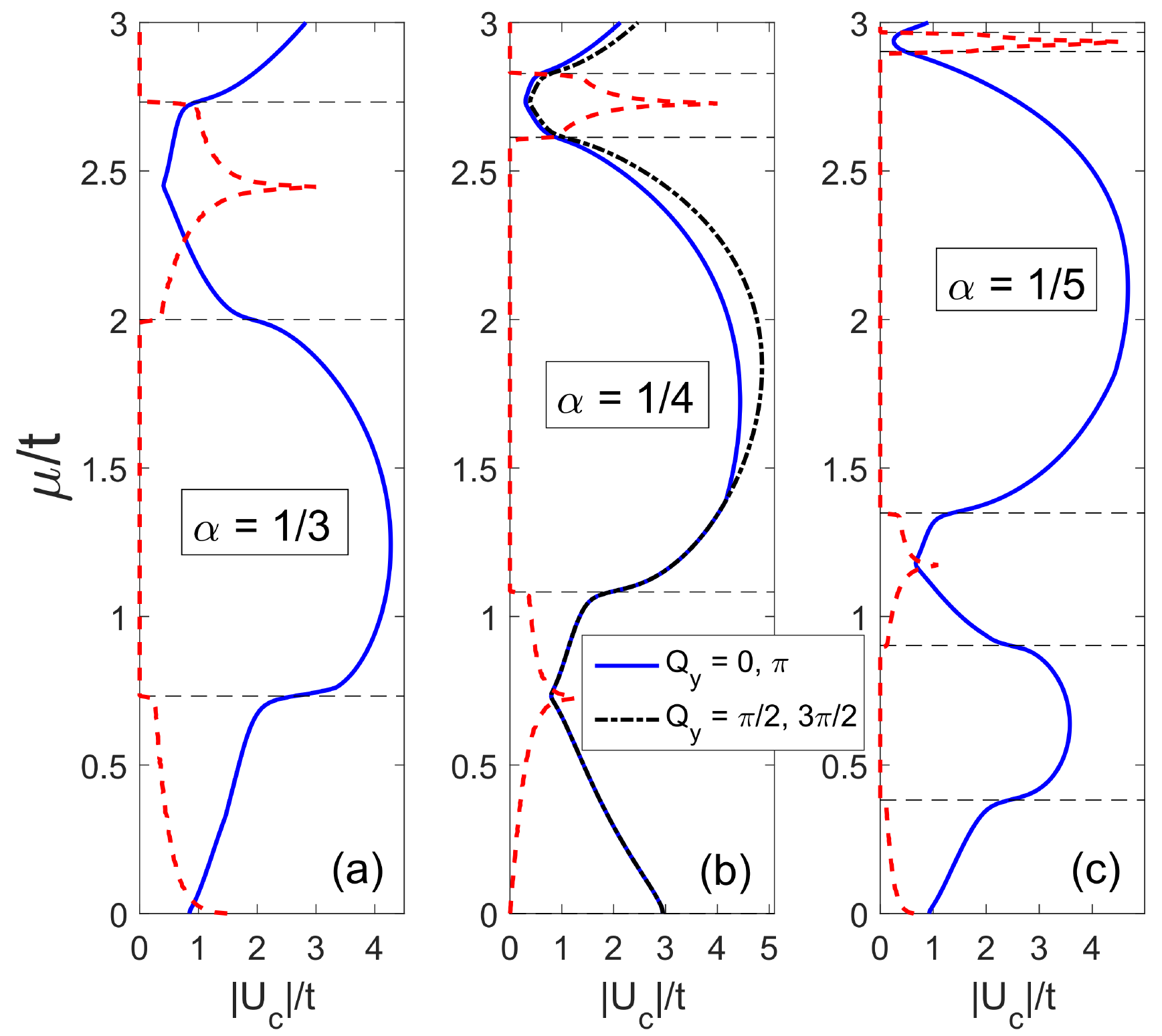}
\caption{(Color online) 
Critical interaction strength $|U_c|/t$ versus the chemical potential $\mu/t$ for 
(a) $\alpha = 1/3$, 
(b) $\alpha = 1/4$, and 
(c) $\alpha = 1/5$. 
Here, $h = 0$ and $k_BT = 0.005t$. 
Density of states $D(\varepsilon)$ is also shown in arbitrary units 
by a red dashed curve. 
Horizontal dashed lines mark the band edges including $\varepsilon = 0$ in (b). 
\label{p1q345h0gcvsmu}
}
\end{figure}

The pairing equation~\eqref{eq:Gap Equation} can also be used to 
determine $|U_c|$ as a function of $\mu$. 
For instance, we consider a population-balanced system with three
distinct $\alpha = 1/3,1/4$ and $1/5$ values in Fig.~\ref{p1q345h0gcvsmu}, 
where we set $h = 0$ for simplicity. 
The results are symmetric around $\mu = 0$ due to the particle-hole 
symmetry of the Hamiltonian.
In all three cases, we observe that the local minima of $|U_c|$ 
coincide intuitively with the local maxima of $D(\varepsilon)$. 
This follows from the fact that when $\mu_\uparrow = \mu_\downarrow$ 
is inside a band and there exist a large number of states in the 
vicinity of $\mu$ available for pairing then $|U_c|$ is small. 
However, as $D(\varepsilon)$ vanishes when $\mu$ enters a band gap, 
$|U_c|$ gets larger attaining its maximum value roughly in the middle 
of the gap. It is remarkable that the maximum values of $|U_c|$ in the 
topmost band gaps of the spectra range somewhere between $4.3t$ 
and $4.9t$ without much variation. 

\begin{figure}[htbp]
\includegraphics[scale=0.45]{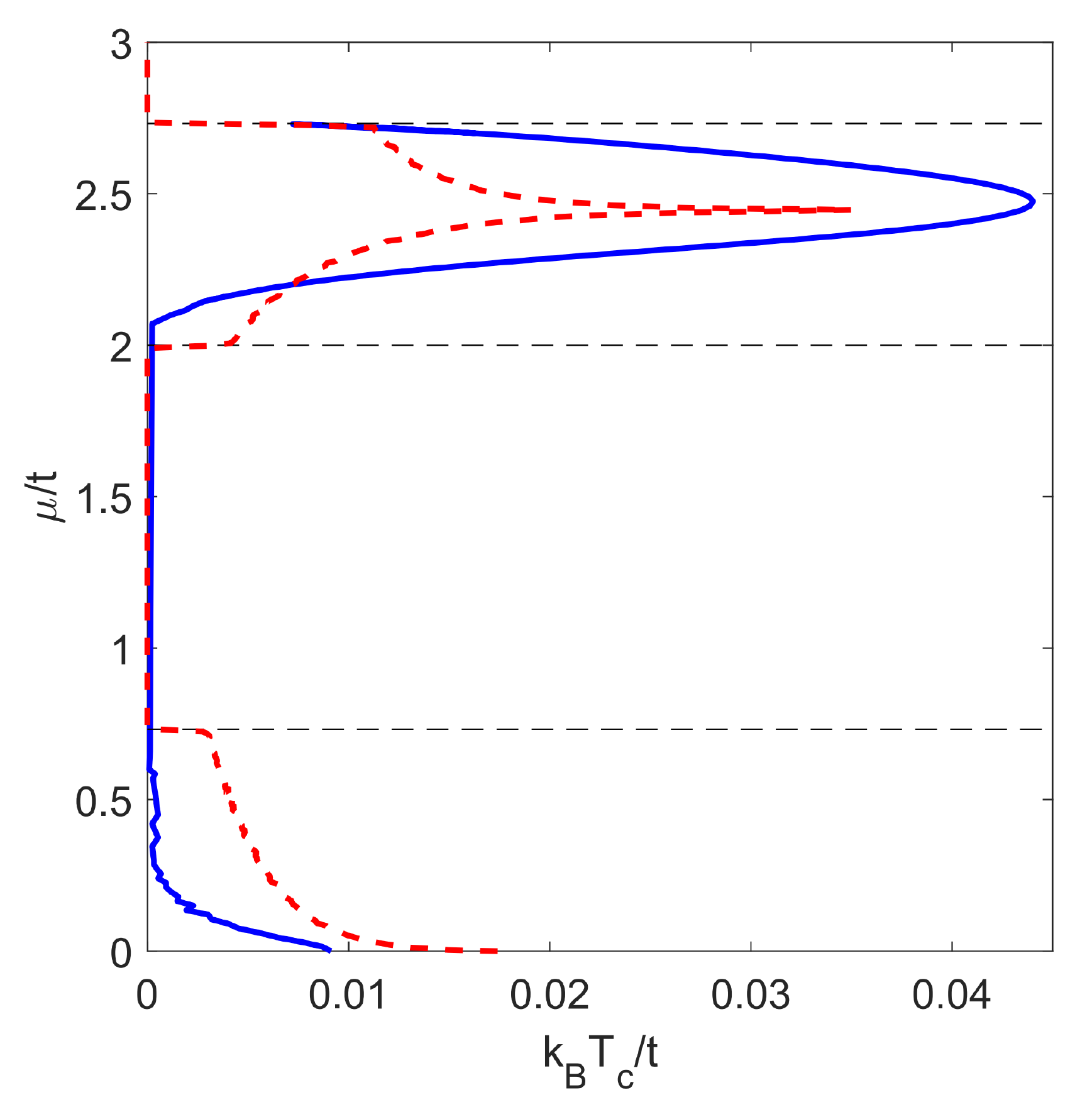}
\caption{(Color online) 
Critical temperature $k_BT_c/t$ as a function of the chemical potential $\mu/t$.
Here, $\alpha = 1/3$, $|U| = t$ and $h=0$. 
Density of states $D(\varepsilon)$ is also shown in arbitrary units 
by a red dashed curve. 
Horizontal dashed lines mark the band edges. 
\label{p1q3Tcvsmu}
}
\end{figure}

As a last application of Eq.~\eqref{eq:Gap Equation}, we determine 
$T_c$ in Fig.~\ref{p1q3Tcvsmu} as a function of $\mu$, where we 
consider $\alpha = 1/3$, and set $|U| = t$ and $h=0$. 
As increasing $T$ weakens the SF state by breaking the Cooper 
pairs, the enhanced pairing due to high $D(\varepsilon)$ is eventually 
beaten by higher $T$. When $\mu$ is inside the band gap, the system 
remains as an insulator even at $T \sim 0$ for the chosen small value 
of $|U|$. However, when $\mu$ is inside a band, the system remains 
as a SF up to a critical $T_c$, the peak values of which coincide with 
the peak values of $D(\varepsilon)$ around the middles of the bands.

\section{Conclusion}
\label{sec:conc}

To summarize, here we analyzed thoroughly the superfluid transition 
in the attractive Hofstadter-Hubbard model, as an attempt to describe 
neutral Fermi gases that are loaded onto square optical lattices and 
subjected to perpendicular and uniform artificial magnetic fields. 
Adopting a BCS-like mean-field approach in momentum space,
we derived a generalized pairing equation in the vicinity of the 
superfluid transition. We solved this equation for the critical interaction 
strength and critical temperature as functions of the Zeeman field and 
chemical potential, by taking primarily into account the finite center-of-mass 
momentum pairing caused by the degeneracies of the single-particle 
Hofstadter spectrum. The non-monotonic variations of the critical 
interaction strength and critical temperature are traced back to the 
sharp changes in the density of single-particle states and to the 
multiple bands of the Hofstadter spectrum, justifying the re-entrant 
superfluidity behavior found in the phase diagrams. 

An extension of this study would be to determine more precisely the 
contribution of Cooper pairs with additional center-of-mass momentum 
in building up especially the superfluid state. This may be accomplished 
through an optimization procedure which fully accounts for the interplay 
between the Zeeman field and the complex band structure arising from 
the artificial magnetic field. As another avenue, our analysis for a square 
lattice can be extended to different lattice geometries like triangular or 
honeycomb ones in light of the recent experiments which demonstrated
the possibility of deforming different lattice types into one another
by tuning lattice parameters~\cite{Tunable}. In particular, it would be 
interesting to study how the topological transitions~\cite{Topological} that 
could be effected in such tunable lattices change the phase boundaries 
under the combined action of a complex band structure and population 
imbalance.

\begin{acknowledgments}
R. O. U. is supported by the T{\"U}B{\.I}TAK B{\.I}DEB 2232 Program and 
M. I. acknowledges funding from T{\"U}B{\.I}TAK Grant No. 1001-114F232 
and the BAGEP award of the Turkish Science Academy. 
\end{acknowledgments}

\appendix
\section{Single-Particle Spectrum}
\label{sec:sps}

\begin{figure}[htbp]
\includegraphics[scale=0.45] {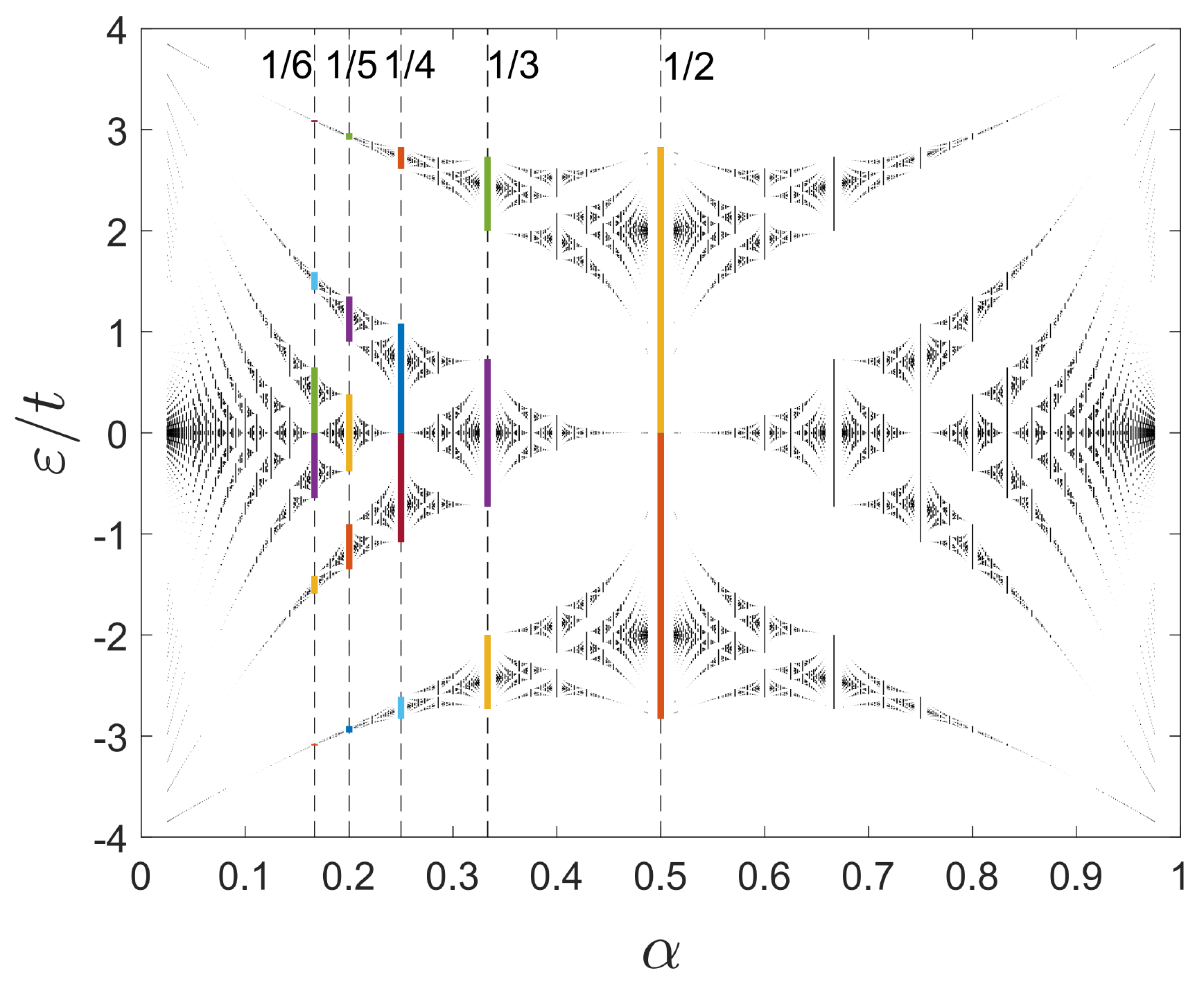}
\caption{(Color online) 
Hofstadter butterfly spectrum displaying the single-particle energy 
$\varepsilon/t$ as a function of the magnetic flux quanta per unit cell 
$\alpha=p/q$. For each $\alpha$, there are a total of $q$ energy bands. 
Vertical dashed lines correspond to $\alpha$ values considered in this work. 
Each band at such an $\alpha$ value is represented by a different color. 
When $q$ is even, two bands touch each other at $\varepsilon = 0$. 
\label{Hofstadter_butterfly}
}
\end{figure}

The single-particle spectrum is determined by diagonalizing the 
$q\times q$ matrix
\bea
\!{\mathbb H}_{{\bf k}\sigma} = {\mathbb H}_{{\bf k}} \! = \!\! \left(\!%
\begin{array}{cccccc}
  D_1 & F & 0 & . & 0 & C \\
  F^{\ast} & D_2 & F & 0 & . & 0 \\
  0 & \ddots & \ddots & \ddots & 0 & . \\
  . & 0 & F^{\ast} & D_m & F & 0  \\
  0 & . & 0 & \ddots & \ddots & \ddots \\
  C^{\ast}& 0 & . & 0 & F^{\ast} & D_q \\
\end{array}\!%
\right),
\label{eq:hofsmatrix}
\eea
where $D_m = -2t\cos(2\pi m\alpha-k_y)$, $F = -t$, and $C = -te^{-i qk_x}$. 
Note that reversing the sign of $\alpha \rightarrow -\alpha$ changes 
the diagonal elements to $D^{\prime}_m = -2t\cos(2\pi m\alpha+k_y)$,
having no effect on our results since such a change corresponds to 
reversing the direction of the magnetic field. In addition, a basis 
transformation of the form 
$
c_{{\bf k}\beta\sigma}\rightarrow c_{{\bf k}\beta\sigma} e^{\pm i\beta k_x}
$ 
changes $F$ to $F^{\prime} = -te^{\pm i k_x}$ and $C$ to 
$C^{\prime} = -te^{\mp i k_x}$, having again no effect on our results.
Representations of the Hofstadter matrix~\eqref{eq:hofsmatrix} with the 
primed quantities are occasionally encountered in the literature~\cite{Kohmoto}. 
In Fig.~\ref{Hofstadter_butterfly}, we show the Hofstadter butterfly 
spectrum as determined by the eigenvalues of the matrix~\eqref{eq:hofsmatrix} 
for all ${\bf k}$ in the first MBZ.

\section{Generalized Pairing Equation}
\label{sec:gpe}

For completeness, here we follow the book by Poole {\it et al.} \cite{Poole}, 
and briefly outline the derivation of Eq.~\eqref{eq:Gap Equation}.
For an alternative method of derivation see the book by de 
Gennes~\cite{de Gennes}. We work in the interaction picture with 
$\tau = it$ the imaginary time, and treat the interaction 
Hamiltonian~\eqref{eq:interaction} as a perturbation. 

The propagator $\kappa(\tau)$ satisfies
\be 
-\frac{d\kappa(\tau)}{d\tau}=H_I(\tau)\kappa(\tau)
\label{eq:PropagatorDifEq},
\ee
where $H_I(\tau)$ is the interaction Hamiltonian $H_I$ in the interaction 
picture 
\be 
H_I(\tau)=e^{\tau H_0}H_Ie^{-\tau H_0}.
\label{eq:Hamiltonian Int pic}
\ee
By integrating Eq.~\eqref{eq:PropagatorDifEq} to first order in $H_I$, we find
\be \kappa(\tau) \simeq 1-\int_0^{\tau}H_I(\tau^{\prime})d\tau^{\prime}.\label{eq:Propagator}
\ee
The inverse of the propagator in this approximation
\be 
\kappa^{-1}(\tau) \simeq 1+\int_0^{\tau}H_I(\tau^{\prime})d\tau^{\prime}
\label{eq:InvPropagator}
\ee
obeys the relation $\kappa^{-1}(\tau)\kappa(\tau)=1$ up to first order 
in $\Delta^l_{\beta}$.

In order to construct a self-consistent equation for $\Delta^l_{\beta}$, we need to determine
the average value 
$
\langle d_{{\bf k}^{l}_{-}n^\prime\downarrow}d_{{\bf k}^{l}_{+}n\uparrow}\rangle
$ 
in terms of $\Delta^l_{\beta}$. For this purpose, using 
$
\langle\ldots\rangle = Z^{-1}{\rm Tr}[e^{-\tau H}\ldots]
$ 
with $Z$ the partition function, and the cyclic property of the trace, 
we find
\bea \langle d_{{\bf k}^{l}_{-}n^{\prime}\downarrow}
d_{{\bf k}^{l}_{+}n\uparrow}\rangle \!=\!\langle 
d_{{\bf k}^{l}_{+}n\uparrow}(\tau)
d_{{\bf k}^{l}_{-}n^{\prime}\downarrow}\rangle, 
\label{eq:dd_average}
\eea
where we define
\bea d_{{\bf k}^{l}_{+}n\uparrow}(\tau) 
&\equiv& e^{\tau H}d_{{\bf k}^{l}_{+}n\uparrow}e^{-\tau H}\nonumber\\
& = & e^{-\tau\epsilon_{{\bf k}^{l}_{+}n\uparrow}}\kappa^{-1}(\tau)
d_{{\bf k}^{l}_{+}n\uparrow}\kappa(\tau).
\label{eq:d_tau}
\eea
An explicit form for $d_{{\bf k}^{l}_{+}n\uparrow}(\tau)$ can be found by 
using Eqs.~\eqref{eq:Hamiltonian Int pic} -~\eqref{eq:InvPropagator} 
and~\eqref{eq:d_tau}, and keeping terms up to first order in 
$\Delta^l_{\beta}$, leading to
\bea d_{{\bf k}^{l}_{+}n\uparrow}(\tau)=&&\!\!\!\!\!\!
e^{-\tau\epsilon_{{\bf k}^{l}_{+}n\uparrow}}
\left[d_{{\bf k}^{l}_{+}n\uparrow}\right.\nonumber\\
&+&\left.\sum\limits_{n^{\prime}l^{\prime}\beta^{\prime}}
\Delta^{l^{\prime}}_{\beta^{\prime}}g^{n *}_{\beta^\prime}({\bf k}^{l}_{+})
g^{n^\prime *}_{\beta^\prime}({\bf k}^{l}_{-}+{\bf Q}_{l^{\prime}})
\right.\label{eq:d_tau explicit} \\
&\times&\left.\frac{e^{\tau(\epsilon_{{\bf k}^{l}_{+}n\uparrow}+\epsilon_{-{\bf k}^{l}_{+}+{\bf Q}_{l^{\prime}},n^{\prime}\downarrow})}-1}{\epsilon_{{\bf k}^{l}_{+}n\uparrow}+\epsilon_{-{\bf k}^{l}_{+}+{\bf Q}_{l^{\prime}},n^{\prime}\downarrow}}d^{\dagger}_{-{\bf k}^{l}_{+}+{\bf Q}_{l^{\prime}},n^{\prime}\downarrow}\right].
\nonumber
\eea
Then, inserting Eq.~(\ref{eq:d_tau explicit}) into Eq.~\eqref{eq:dd_average}, 
we find
\bea 
\langle d_{{\bf k}^{l}_{-}n^{\prime}\downarrow}
d_{{\bf k}^{l}_{+}n\uparrow}\rangle &=&\sum\limits_{\beta^{\prime}}
\Delta^{l}_{\beta^\prime}g^{n *}_{\beta^\prime}({\bf k}^{l}_{+})
g^{n^\prime *}_{\beta^\prime}({\bf k}^{l}_{-}) 
f(\epsilon_{{\bf k}^{l}_{-}n^\prime\downarrow})
\nonumber\\ 
&\times& 
f(\epsilon_{{\bf k}^{l}_{+}n\uparrow})
\frac{e^{\tau(\epsilon_{{\bf k}^{l}_{+}n\uparrow}+\epsilon_{{\bf k}^{l}_{-}n^\prime\downarrow})}-1}
{\epsilon_{{\bf k}^{l}_{+}n\uparrow}+\epsilon_{{\bf k}^{l}_{-}n^\prime\downarrow}},
\label{eq:dd_average final}
\eea
where $f(x) = 1/(e^{\tau x} + 1)$. Here, we use 
$
\langle d^{\dagger}_{\alpha}d_{\gamma}\rangle \simeq \delta_{\alpha\gamma}f(\epsilon_{\alpha}),
$ 
which is valid up to first order. Finally, inserting Eq.~\eqref{eq:dd_average final}
into the definition of $\Delta^{l}_\beta$, rearranging the exponential terms, and replacing $\tau$ with $1/(k_BT)$ we obtain the generalized pairing 
Eq.~\eqref{eq:Gap Equation} given in the main text.

\begin{figure}[htbp]
\includegraphics[scale=0.45] {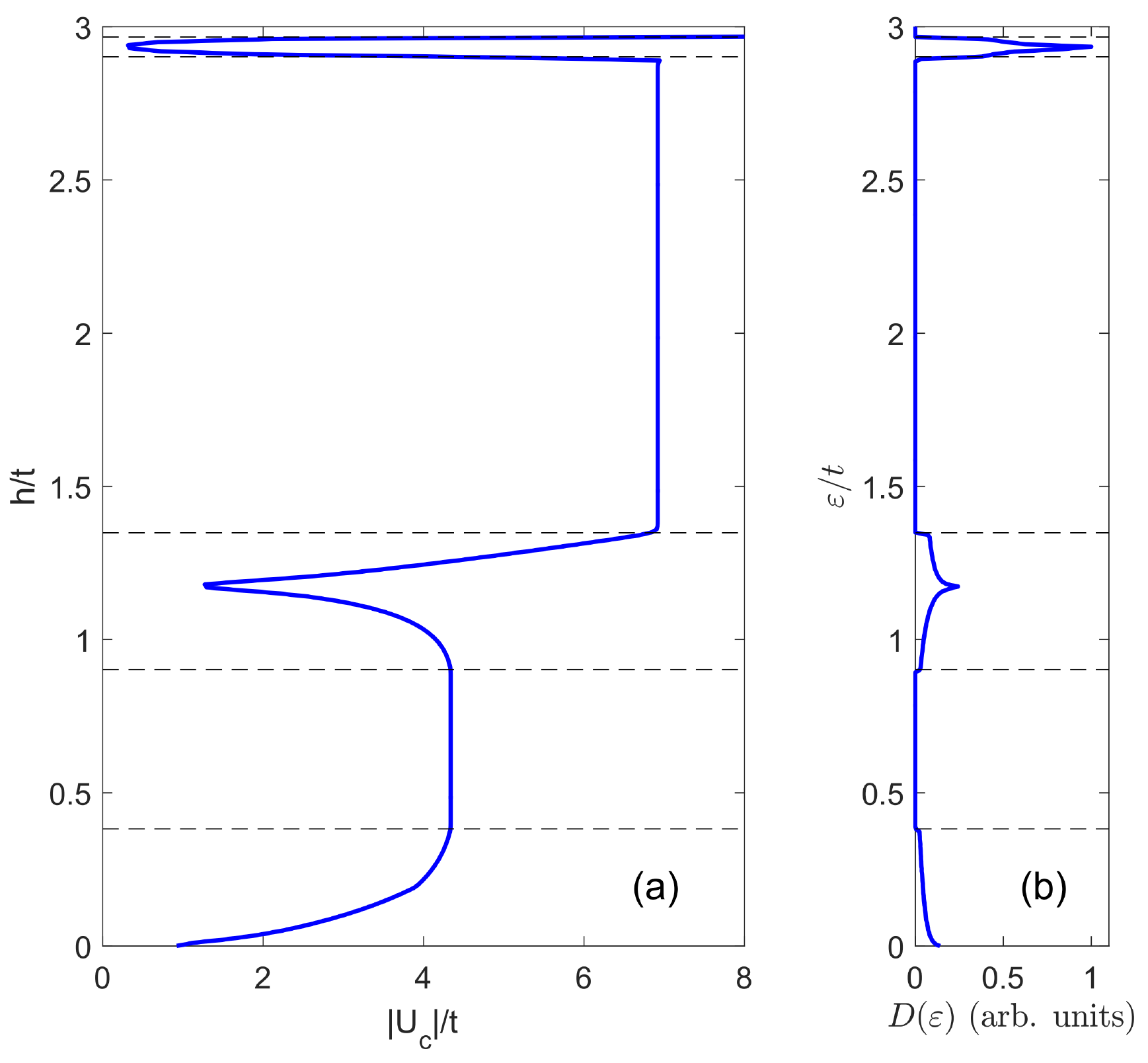}
\caption{(Color online) 
(a) Critical interaction strength $|U_c|/t$ versus the Zeeman field $h/t$ 
with $\alpha = 1/5$, $\mu = 0$ and $k_BT = 0.005t$. Phase boundary 
is degenerate for all ${\bf Q}_l = (0, 2\pi l/5)$ with $l = 0,1,\ldots,4$. 
(b) Density of states $D(\varepsilon)$ in arbitrary units. 
Horizontal dashed lines show the band edges. 
\label{p1q5mu0gcvsh}
}
\end{figure}
\section{Phase Diagrams for $\alpha = 1/5$ and $\alpha = 1/6$}
\label{sec:pd}

\begin{figure}[htbp]
\includegraphics[scale=0.45]{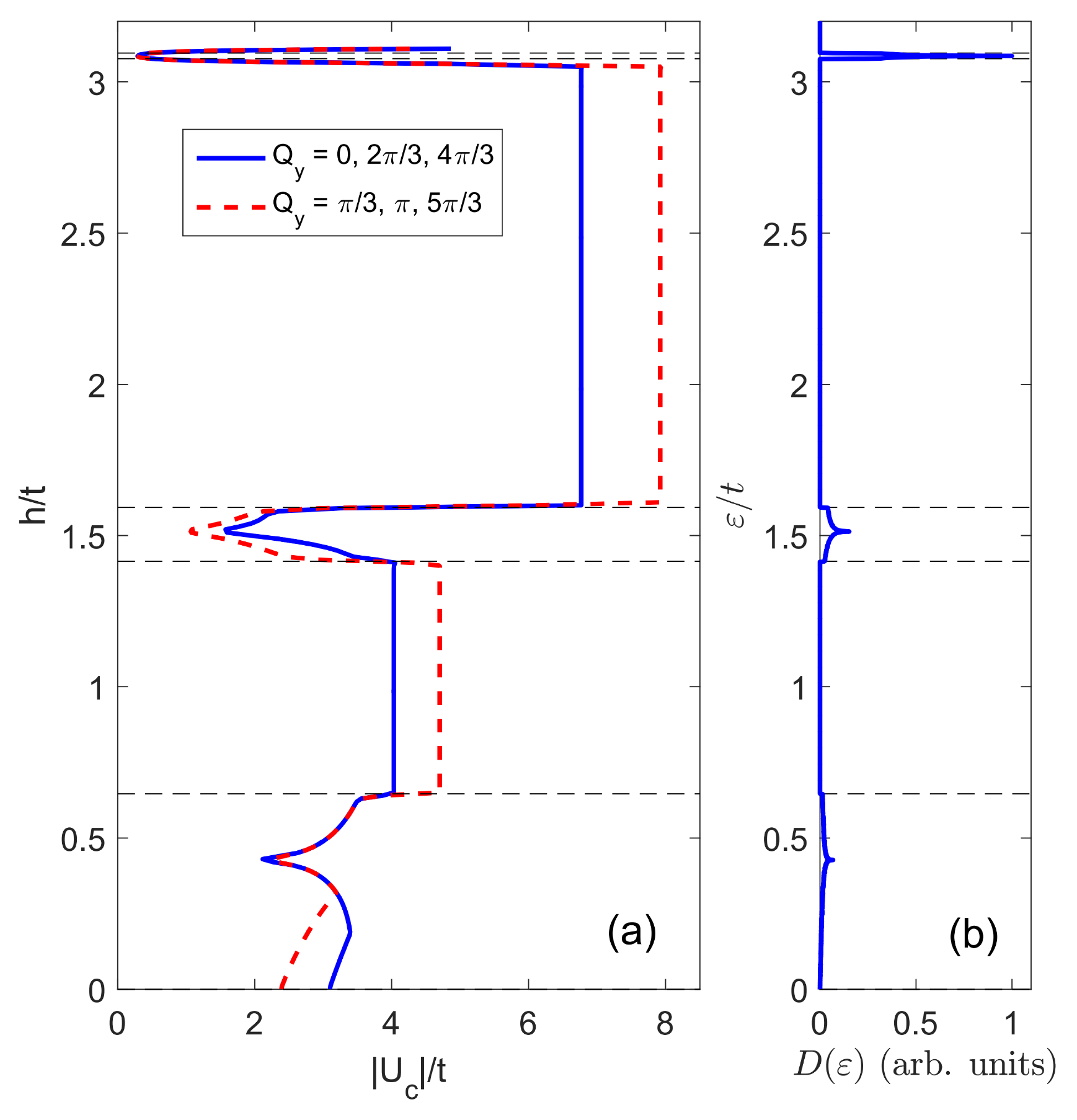}
\caption{(Color online)
(a) Critical interaction strength $|U_c|/t$ versus the Zeeman field $h/t$ 
with $\alpha = 1/6$, $\mu = 0$ and $k_BT = 0.005t$. 
Solid blue curve is for ${\bf Q} = (0,0)$, ${\bf Q} = (0,2\pi/3)$, and 
${\bf Q} = (0,4\pi/3)$; red dashed curve is for ${\bf Q} = (0,\pi/3)$, 
${\bf Q} = (0,\pi)$, and ${\bf Q} = (0,5\pi/3)$. 
(b) Density of states $D(\varepsilon)$ in arbitrary units. 
Horizontal dashed lines show the band edges including $\varepsilon = 0$.
\label{p1q6mu0gcvsh}
}
\end{figure}

For these lower magnetic flux values, since there are, respectively, 
five and six subbands in the energy spectrum, Figs.~\ref{p1q5mu0gcvsh} 
and~\ref{p1q6mu0gcvsh} show much narrower bands in comparison
to those presented in the main text with smaller $q$. 
In particular, the highest bands (as well as the lowest ones which 
are not shown) of the $\alpha=1/5$ and $1/6$ spectra turn out to 
be very narrow, causing a sharp variation of $D(\varepsilon)$ with 
$\varepsilon$ and giving rise to a large dip in $|U_c|$ as a function 
of increasing $h$. A notable distinction between these two cases is 
that while $U_c^l$ are degenerate functions of $h$ for all 
${\bf Q}_l=(0, 2\pi l p/q)$ with $l = 0,1,\ldots,4$ when $\alpha = 1/5$, 
there are two distinct solution sets corresponding to $l=(0,2,4)$ and 
$l=(1,3,5)$ when $\alpha=1/6$. 

As a final remark, we note that all of our numerical results for low 
$q = 1,2,\ldots,6$ values show that while $U_c^l$ are degenerate 
functions of $h$ for all ${\bf Q}_l$ when $q$ is odd, there are two 
distinct solution sets corresponding to $l = (0,2,\ldots,q-2)$ and 
$l = (1,3,\ldots,q-1)$ when $q$ is even. It would be curious to check 
whether this observation applies to arbitrary $q$ values in general.

\end{document}